\documentstyle[lprocldpf,psfig,11pt]{article}

\def\Journal#1#2#3#4{{#1} {\bf #2}, #3 (#4)}

\def\NPA{{\em Nucl. Phys.} A}

\def\PLB{{\em Phys. Lett.} B}
\def\PRP{\em Phys. Rep.}
\def\PRL{\em Phys. Rev. Lett.}
\def\PRC{{\em Phys. Rev.} C}

\def\PPN{\em Prog. Part. Nucl. Phys.}

\def\mco{\multicolumn}

\def\vep{\varepsilon}

\def\be{\begin{equation}}
\def\ee{\end{equation}}
\def\bea{\begin{eqnarray}}
\def\eea{\end{eqnarray}}

\begin{document}

\title{Double Giant Dipole Resonance in $^{208}$Pb}


\author{ S. NISHIZAKI }

\address{Faculty of Humanities and Social Sciences, Iwate University,\\
3-18-34 Ueda, Morioka 020, Japan}

\author{ J. Wambach }

\address{Institut f\"{u}r Kernphysik, Technische Hochschule Darmstadt,\\
Schlossgartenstrasse 9, D-64289 Darmstadt, Germany}


\maketitle\abstracts{
Double-dipole excitations in $^{208}$Pb are analyzed within a microscopic
model explicitly treating 2p2h-excitations. Collective states
built from such 2p2h-excitations are shown to appear at about twice 
the energy of the isovector giant dipole resonance, in agreement with
the experimental findings. The calculated cross section for Coulomb 
excitation at relativistic energies cannot explain simultaneously
the measured single-dipole and double-dipole cross sections, however.}

\section{Introduction}
Double giant dipole resonances (DGDR) have been observed
in pion double-charge-exchange reactions at LAMPF \cite{Mor}
as well as in high-energy heavy-ion collisions at GSI \cite{Rit}.
In the latter case there is a strong focusing of the electromagnetic
field in the target rest frame. This greatly enhances the field
intensity in the vicinity of the target nucleus, thus increasing
the probability for two-photon absorption from the ground state.

The global parameters of the DGDR, {\sl i.e.} the resonance energy 
$E_{DGDR}$, the width $\Gamma_{DGDR}$ and
the cross section $\sigma$ can be summarized as follows \cite{Eml}:
1) The resonance energy is about twice as large as that of the isovector
giant dipole resonance (GDR), $E_{DGDR}\simeq 2 E_{GDR}$.
which suggests that can be interpreted as an independent two-phonon state.
2) The observed values of the width $\Gamma_{DGDR}$ are bracketed by
  $\sqrt{2}\Gamma_{GDR}$ and $2\Gamma_{GDR}$.
Also this feature supports a harmonic picture.
3) The measured cross section $\sigma_{{\rm exp}}$ is larger than
the theoretical estimate $\sigma_{{\rm th}}$. The values of the ratio 
$\sigma_{{\rm exp}}/\sigma_{{\rm th}}$ are scattered between 1 and 5
depending on the nuclei considered and on the theoretical analyses.
This puts into some doubt to interpret the DGDR as an independent 
two-phonon state.

To reach a quantitative understanding of the DGDR characteristics
we shall carry out a microscopic calculation of the DGDR in $^{208}$Pb.
The structure of double-dipole states are investigated by the 
diagonalization of a model Hamiltonian in the space of 1p1h- and 
2p2h-excitations. For comparison with experiment the cross section for 
Coulomb excitation, based on the second-order perturbation theory, will be
estimated.
  
\section{Double-Dipole Strength Distribution in $^{208}$Pb}
\subsection{Microscopic Model}

The nuclear structure model has been developed in our previous
work \cite{Nis} on double-dipole excitations in $^{40}$Ca. 
We give a brief outline stressing some new aspects in
the application to heavy nuclei such as $^{208}$Pb.
The eigenstates $|n\rangle$ of the nuclear Hamiltonian $\hat H_0$ 
are determined by diagonalizing $\hat H_0$ in the space of 1p1h- and 
2p2h-excitations. As a model space of single-particle states, we 
take into account three major shells of the Woods-Saxon potential
on both sides of the Fermi level. The continuum states are
treated by discretization through an expansion in a harmonic
oscillator basis. The number of 2p2h states
constructed with these single-particle states exceeds several
hundreds of thousand. Most of them, however, have vanishingly small
matrix elements for the double-dipole transition from the ground
state. Thus we select only those 2p2h-states whose energy is below 40 MeV, 
and which exceed a lower limit in the magnitude of the double-dipole matrix 
element,
\be  
 |\langle 2p2h|{\hat D}{\hat D}|0\rangle|^2/
 \sum_{2}|\langle 2|{\hat D}{\hat D}|0\rangle|^2
 \geq 5\times 10^{-5}.
\label{eq:sel}
\ee
\noindent
Here, the isovector dipole operator, ${\hat D}$, is defined as
\be
 {\hat D}=e\frac{N}{A}\sum_{i=1}^{Z}{\bf r}_i
         -e\frac{Z}{A}\sum_{i=1}^{N}{\bf r}_i .
\label{eq:ope}
\ee
\noindent
With these conditions, we can reduce the number of
2p2h-states to 996 for the $J^{\pi}=0^+$ double-dipole transition
and to 2011 for the $2^+$ transition.  
 
The mean energy and total strength of the GDR in $^{208}$Pb,
calculated in the usual 1p1h-RPA with the nuclear Hamiltonian \cite{Nis},
including a density-dependent zero-range interaction, 
are 10.3 MeV and 98\% of the TRK sum-rule value, respectively. Both are 
smaller than photo-neutron data \cite{Vey} which yield 13.4 MeV and 134\%.
Therefore we enhance the theoretical results by a scaling
of the single-particle energies according to $\vep_{sp}=\vep_{WS}/(m^*/m)$
with the effective mass $m^*/m=0.75$. By this procedure,
we obtain the mean energy and the total strength as 13.5MeV
and 131\% of the TRK value, in good agreement with experiment.

\subsection{Strength Distribution}
\begin{figure}[ht]
\centerline{\psfig{figure=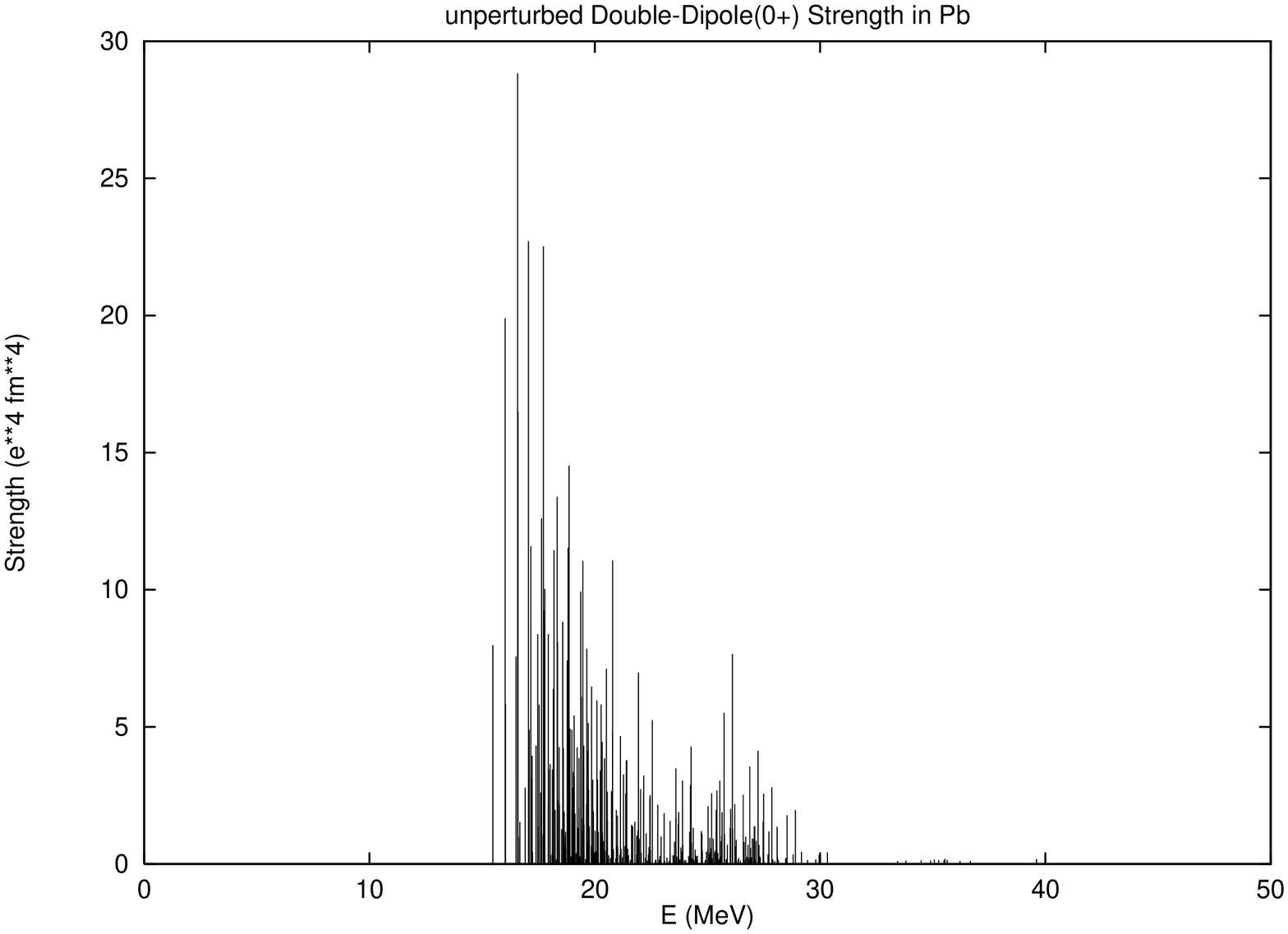,height=5.0cm,angle=270}
\psfig{figure=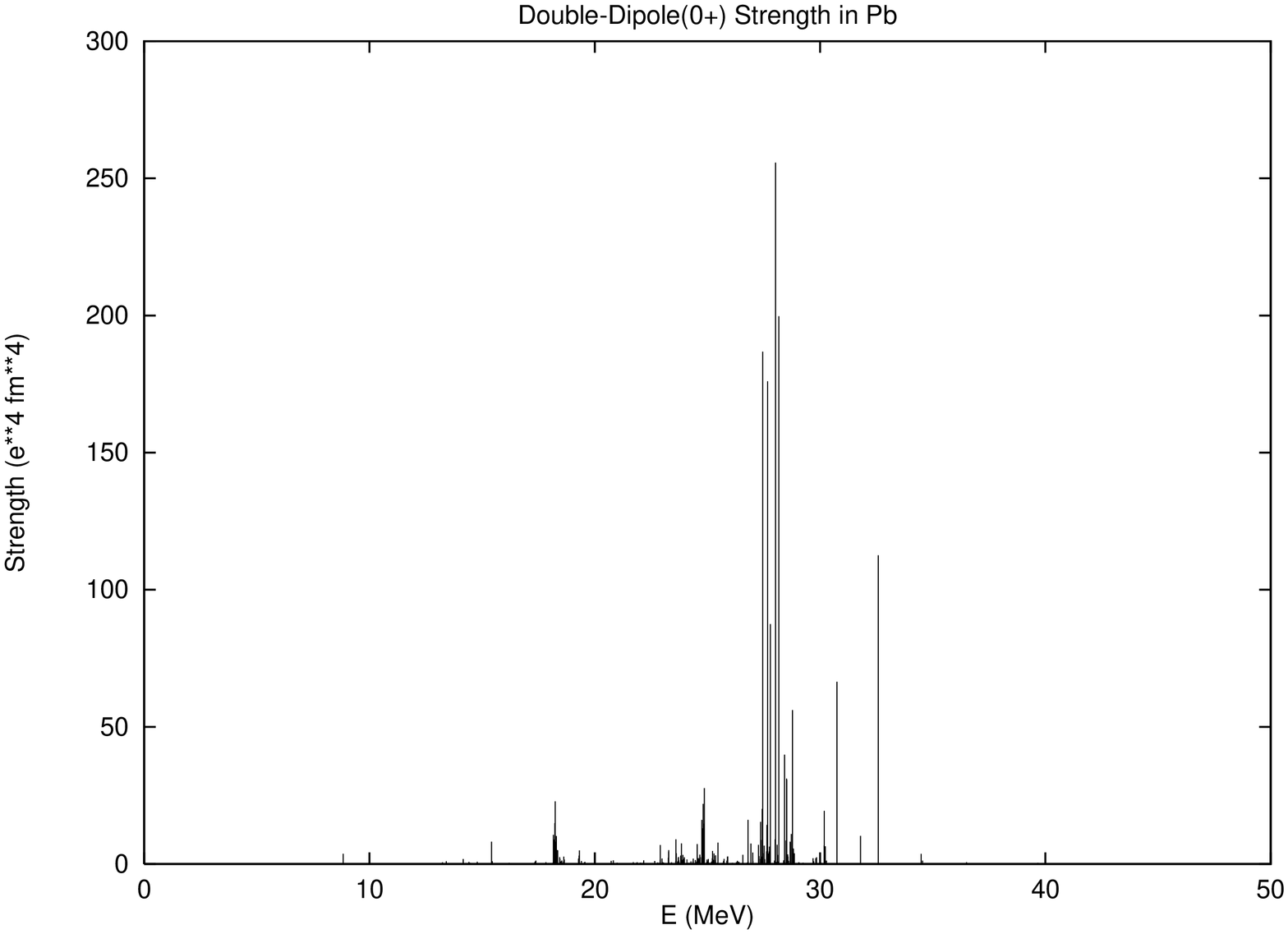,height=5.0cm,angle=270}}
\vskip 0.5cm
\centerline{\psfig{figure=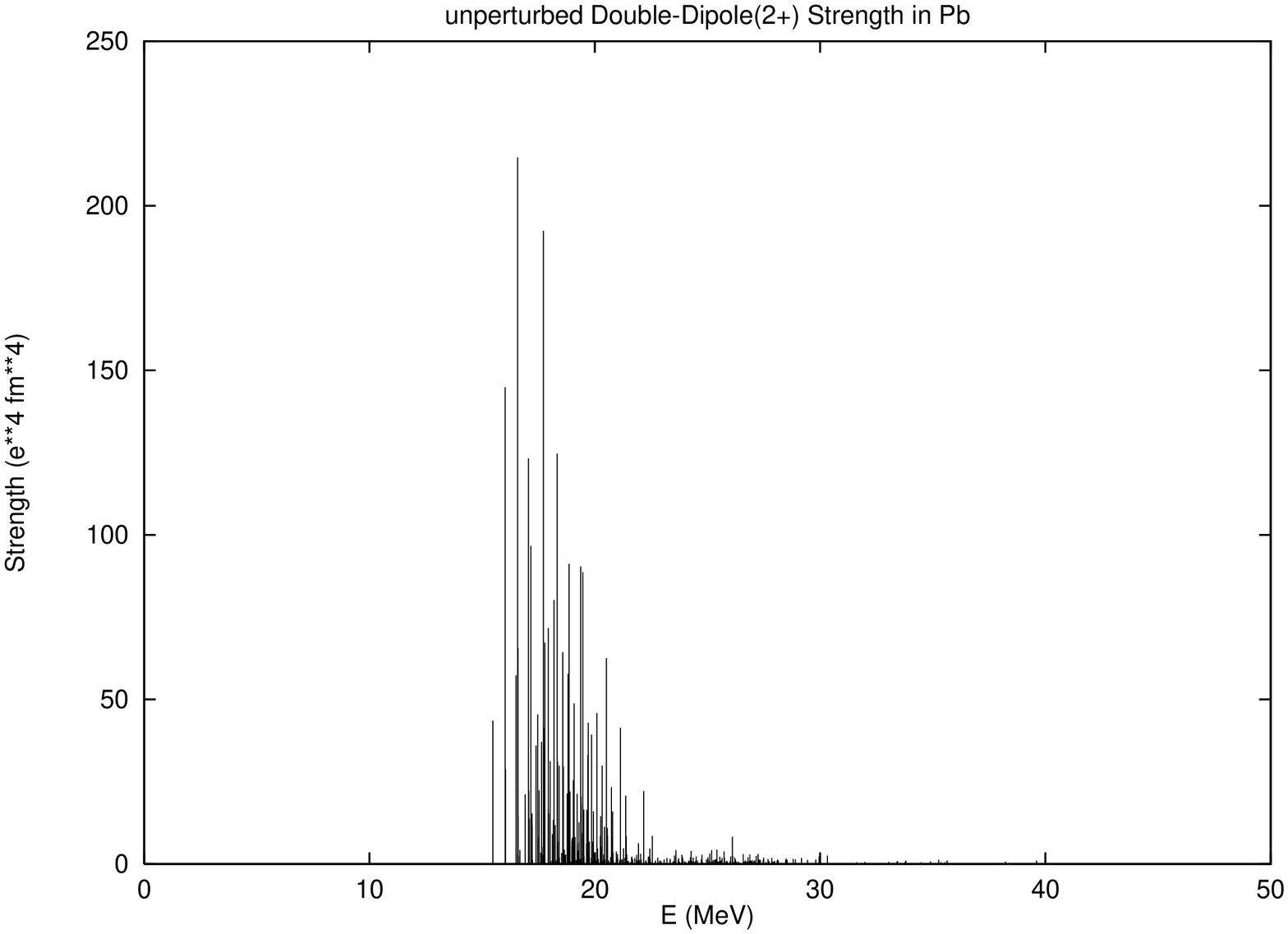,height=5.0cm,angle=270}
\psfig{figure=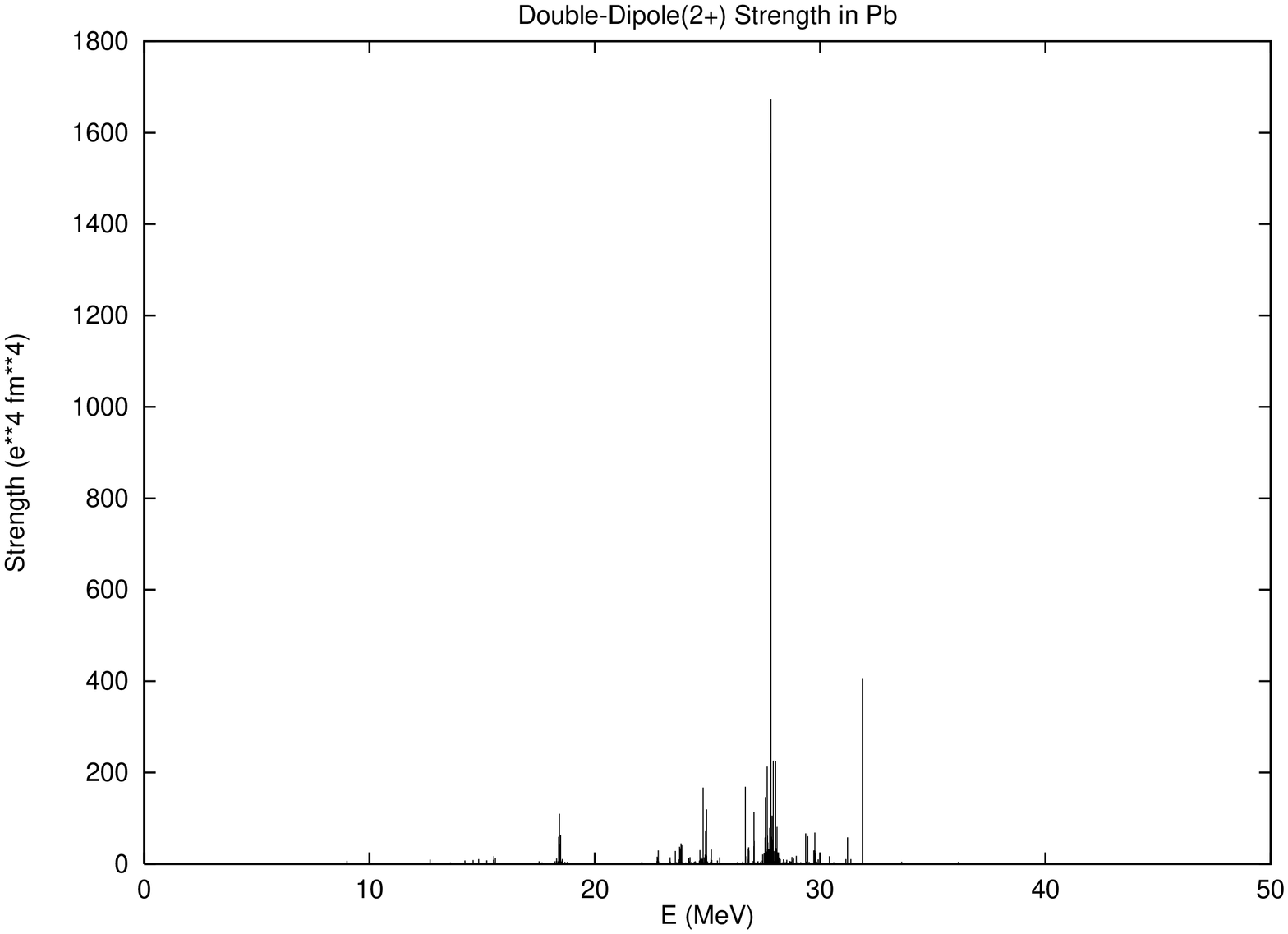,height=5.0cm,angle=270}}
\caption{Double-Dipole Strength Distributions. 
    (a) Unperturbed $0^+$ DD strength.
    (b) Perturbed $0^+$ DD strength.
    (c) Unperturbed $2^+$ DD strength.
    (d) Perturbed $2^+$ DD strength.
\label{fig:str}}
\end{figure}

The strength distributions of the double-dipole transition
are shown in Fig.~\ref{fig:str}. Figs.~1(a) and  1(b) (1(c) and 1(d))
represent the unperturbed and the perturbed strength distributions for the 
$0^+$ ($2^+$) double-dipole transitions, respectively. 
Both for $0^+$ and $2^+$ double-dipole transitions,
the perturbed strength concentrates in the region of 25--30 MeV.
There appear several states which carry about ten times more strength
than the strongest unperturbed ones.

The mean energy, $\langle E\rangle$, and the width, $\Gamma$, are
estimated in terms of the energy moments
\be
 m_{DD}^k=\int dE~E^kS_{DD}(E), \qquad
 S_{DD} = \sum_n
     |\langle n|{\hat D}{\hat D}|0\rangle |^2\delta(E-E_n) ,
\ee
\noindent
as
\be
 \langle E\rangle_{DGDR}=m_{DD}^1/m_{DD}^0 \quad {\rm and} \quad
 \Gamma_{DGDR}=\sqrt{m_{DD}^2m_{DD}^0-(m_{DD}^1)^2}/m_{DD}^0 ,
\ee
\noindent
and similarly for the isovector dipole excitations.
The $\langle E\rangle_{DGDR}$ and $\Gamma_{DGDR}$ are
26.96 MeV (26.53 MeV) and 3.48 MeV (3.36 MeV)
for $0^+$ DGDR ($2^+$ DGDR), respectively.
The mean energy of the DGDR is about twice as large as that of the GDR.
The calculated width, which corresponds to the fragmentation width 
(Landau damping) is larger than that of the GDR ($\Gamma_{GDR}=2.70$ MeV) 
by the factor of 1.24--1.29. We note that the observed width includes, 
in addition, the escaping and spreading width which are not included
in the present calculation.

\section{Cross Section for the Coulomb Excitation}

\subsection{Second-order Coulomb-Excitation Amplitude}

In the semiclassical approach \cite{Ald,Ber}, the cross section for
Coulomb excitation is given by
\be
  \sigma_{i \rightarrow f}=2\pi\int_{{\bf R}_{min}}^{\infty}b~db
 ~\frac{1}{2J_i+1}\sum_{M_i,M_f}|a_{fi}|^2.
\ee
\noindent
The projectile is assumed to move on a straight-line trajectory
with impact parameter $b$. The minimum impact parameter $R_{min}$
is a key parameter, since the cross section sensitively depends on it
as will be shown later.

The main components of the DGDR are 2p2h-states, which are excited via
a two-body operator of double-dipole character ($\hat D\hat D$) from the 
ground state. For Coulomb excitation the precise form of the excitation 
operator is derived from the 2nd-order amplitude 
\be
 a_{fi}^{(2)}=\frac{1}{(i\hbar)^2}\langle f|\int_{-\infty}^{+\infty}dt
 \int_{-\infty}^{t}dt'~e^{+i\hat H_0t/\hbar}\hat V(t)e^{-i\hat H_0t/\hbar}
 e^{+i\hat H_0t'/\hbar}\hat V(t')e^{-i\hat H_0t'/\hbar}|i\rangle ,
\label{eq:amp0}
\ee
\noindent
where $\hat H_0$ and $\hat V$ are the nuclear Hamiltonian and 
the Coulomb interaction
between the projectile and target nuclei, respectively.
To obtain the two-body operator, we exchange the order of 
$\hat V(t)$ ($\hat V(t')$) 
and $\exp{(+i\hat H_0t'/\hbar)}$ ($\exp{(-i\hat H_0t/\hbar)}$). This yields a
series expansion in the times t and t' which includes single- and 
multiple-commutators of $\hat H_0$ and $\hat V$. In the limit of a 'fast collision' 
\cite{Type}, we need only to take into account the lowest order in time. 
Finally the expression for the second-order amplitude of the double 
$E\lambda'$ transition is given by
\bea
 a_{fi}^{(2)}&=&(4\pi\frac{Ze}{i\hbar})^2\sum_{\mu}\frac{(-)^{\mu}}{(2\lambda'+1)^2}
\nonumber\\
             &&\times\{ \frac{1}{2}\langle f|{\cal N}_{EE}^{\lambda'\lambda'}(\lambda,-\mu)
              |i\rangle (T_{\lambda\mu}^{E\lambda' E\lambda'}(\omega,\omega)
               + \hbar\omega U_{\lambda\mu}^{E\lambda'E\lambda'}(\omega,\omega)) 
\nonumber\\
             &&\quad + \langle f|{\cal K}_{EE}^{\lambda'\lambda'}(\lambda,-\mu)|i\rangle
                \frac{i}{2\pi}{\cal P}\int_{-\infty}^{+\infty}\frac{dq}{q}
                U_{\lambda\mu}^{E\lambda'E\lambda'}(\omega-q,\omega+q)\},
\label{eq:amp}
\eea
\noindent
with

\bea
  T_{\lambda\mu}^{\pi_1\lambda_1 \pi_2\lambda_2}(\omega_1,\omega_2)
      &=& \bigl[ S_{\pi_1\lambda_1}(\omega_1)\otimes                     S_{\pi_2\lambda_2}(\omega_2)\bigr]_{\lambda\mu}\\
  U_{\lambda\mu}^{\pi_1\lambda_1 \pi_2\lambda_2}(\omega_1,\omega_2)
      &=& \bigl[ S_{\pi_1\lambda_1}(\omega_1)\otimes                     R_{\pi_2\lambda_2}(\omega_2)\bigr]_{\lambda\mu}\\
 {\cal N}^{\lambda_1\lambda_2}_{\pi_1\pi_2}(\lambda,\mu)
      &=& \bigl[ {\cal M}(\pi_1\lambda_1)\otimes 
          {\cal M}(\pi_2\lambda_2)\bigr]_{\lambda\mu}\\
 {\cal K}^{\lambda_1\lambda_2}_{\pi_1\pi_2}(\lambda,\mu)
      &=&\bigl[ [ H_0, {\cal M}(\pi_1\lambda_1)] \otimes 
          {\cal M}(\pi_2\lambda_2) \nonumber \\
      && -{\cal M}(\pi_1\lambda_1) \otimes 
           [ H_0, {\cal M}(\pi_2\lambda_2)] \bigr]_{\lambda\mu}.
\eea
\noindent
Here, ${\cal M}(\pi \lambda)$ denotes electric ($\pi=E$) or magnetic 
($\pi=M$) multipole operators while $S_{\pi\lambda}$ and $R_{\pi\lambda}$ 
are orbital integrals \cite{Type}. The principal value integral in 
Eq.~\ref{eq:amp} results from a step function, which appears when the 
upper limit of the integral in Eq.~\ref{eq:amp0} is extended to infinity. 
In the present estimate of the cross section, we neglect this principal 
value integral.

\subsection{Cross Section}
\begin{figure}[ht]
\centerline{\psfig{figure=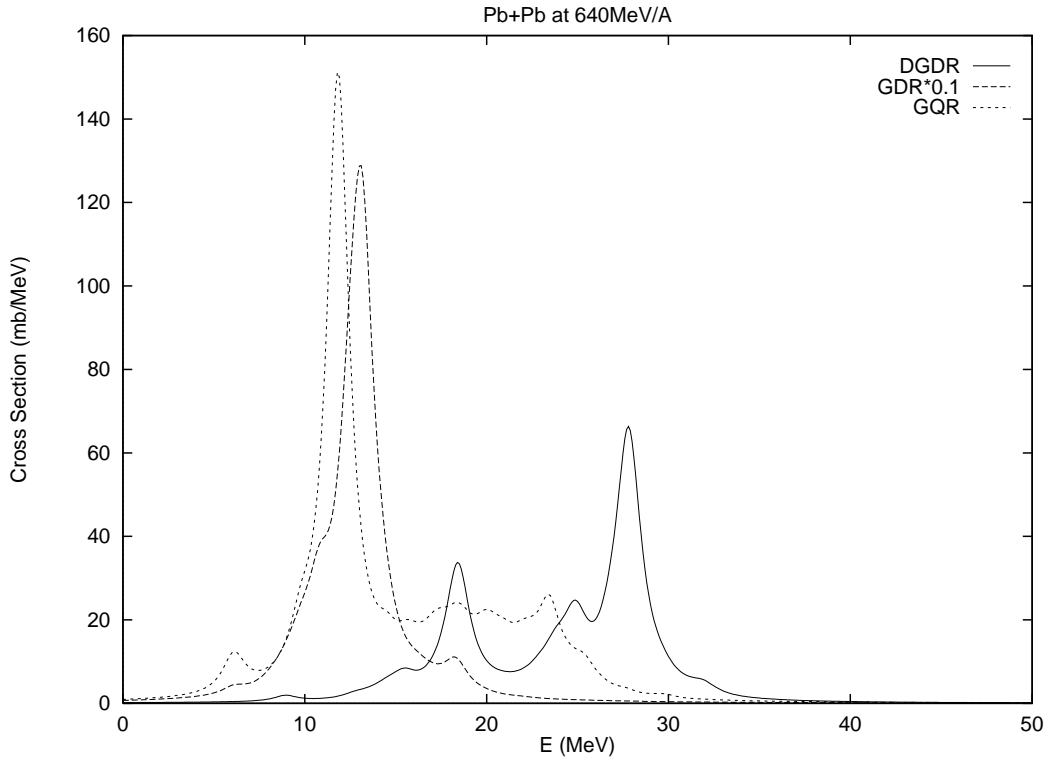,height=10.0cm,angle=270}}
\caption{Cross Section for the Coulomb Excitation. 
The solid line denotes the cross section of the DGDR
while the dashed line displays the GDR cross section (multiplied by 0.1).
In addition, the cross section for the GQR is shown as the dotted line. The 
minimum impact parameter is the value of case (b) (see text).
\label{fig:crs}}
\end{figure}

The calculated cross section for Coulomb excitation of the DGDR with a 
$^{208}$Pb 
projectile incident on a Pb target at 640 MeV/A is plotted as the solid 
line in Fig.~\ref{fig:crs}. The dashed and dotted lines display the cross 
sections for the GDR and the GQR (giant quadrupole resonance), respectively.
For comparison the former has been multiplied by the factor of 0.1. 
To account for experimental energy resolution, the calculated cross 
sections have been smoothed by using a Breit-Wigner function
of width 1.5 MeV. The main peak of the DGDR appears in the region of
25 -- 30 MeV, just above the broad isovector GQR around 20 MeV. The peak 
energy of the DGDR is about twice that of the GDR.

Next we discuss the dependence of the energy-integrated cross section
on the minimum impact-parameter, $R_{min}$. We parameterize $R_{min}$ as 
$R_{min}=r_0(A_t^{1/3}+A_p^{1/3})$, where $A_t$ and $A_p$ denote the mass
number of target and projectile nuclei, respectively. Three cases are 
considered : (a) $r_0=1.50$ fm, (b) $r_0=1.31$ fm and (c) $r_0=1.20$ fm. 
The values for (a) and (c) have been used previously by Ponomarev et 
al.~\cite{Pon} in the study of the DGDR in $^{136}$Xe, while   
case (b) represents the choice of Boretzky et al.~\cite{Bor} 
(Fig.~\ref{fig:crs} corresponds to case (b)). The cross sections of the 
GDR and the DGDR in $^{208}$Pb
incident on various nuclei are given in Table \ref{tbl:crs} 
where the three cases for $R_{min}$ are compared with the measured 
cross sections~\cite{Bor}. The intermediate value of $R_{min}$ (case (b)) 
reproduces the measured cross sections of the DGDR fairly well, but 
overestimates those of the GDR. On the other hand,
the larger value (case (a)) reproduces the measured cross sections of
GDR, while underestimating those of the DGDR by the factor 2 -- 3.
In our calculation it seems impossible to simultaneously explain the 
measured GDR and DGDR cross sections with the same value of $R_{min}$ .

In the particle-vibration coupling calculation of the DGDR in $^{136}$Xe
by Ponomarev et al. \cite{Pon}, similar results for the cross section
were obtained, but the discrepancy between the measured cross section
and their estimate is larger than that in our results. On the other hand,
the simple folding model analysis of the DGDR in $^{208}$Pb by Boretzky 
et al. \cite{Bor} predicts only a 33\% enhancement of the DGDR cross section
while reproducing the GDR cross section. 
It remains an open question whether this difference between results from 
the microscopic models and the folding model originates from 
the nuclear structure or from the treatment of the reaction
mechanism.

\begin{table}[ht]
\caption{Cross section for Coulomb excitation of the GDR and DGDR in 
a $^{208}$Pb projectile incident on U, Pb, Ho and Sn targets at 640 MeV/A.
\label{tab:crs}}
\vspace{0.4cm}
\begin{center}
\begin{tabular}{|c|c|c|c|c|c|c|c|c|}\hline
 Cross&\mco{2}{c|}{U} &\mco{2}{c|}{Pb}
 &\mco{2}{c|}{Ho} &\mco{2}{c|}{Sn} \\
\cline{2-9}
 Section (b)&GDR&DGDR&GDR&DGDR&GDR&DGRD&GDR&DGDR \\ \hline
(a) & 4.18 & 0.222 & 3.47 & 0.162 & 2.49 & 0.091 & 1.52 & 0.038 \\
(b) & 5.34 & 0.500 & 4.42 & 0.361 & 3.16 & 0.201 & 1.91 & 0.083 \\
(c) & 6.22 & 0.827 & 5.12 & 0.590 & 3.63 & 0.321 & 2.18 & 0.129 \\
exp & 3.66 & 0.51  & 3.28 & 0.38  & 2.47 & 0.28  & 1.45 & 0.07  \\ \hline
\end{tabular}
\end{center}
\label{tbl:crs}
\end{table}

\section{Summary}
We have carried out a microscopic calculation of double-dipole excitations 
in $^{208}$Pb in a realistic model space of 1p1h- and 2p2h-states. Since the 
number of 2p2h-states in heavy nuclei such as $^{208}$Pb, is 
prohibitively large we have introduced the selection procedure given by 
Eq.~\ref{eq:sel} as well as a truncation in the 2p2h energy. Although the 
width of DGDR may not be totally reliable because of this procedure, we 
believe that both the mean energy and the integrated strength are realistic. 

The double-dipole strength is shown to be concentrated in
an energy region twice that of the isovector giant dipole resonance 
thus indicating that anharmonicity effects are quite small, in good 
agreement with experiment.

The Coulomb excitation cross section has been
calculated based on the second-order perturbation.
It depends sensitively on the minimum impact parameter, which
appears in the semiclassical treatment of the cross
section for Coulomb excitation. We have estimated the
cross section of the GDR and DGDR with three choices
for the minimum impact parameter, available from the literature.
None of these choices can explain the measured cross sections 
simultaneously and hence the discrepancy between the measured
cross section and theoretical estimates, previously observed, remains.

\section*{Acknowledgments}
We thank G. Baur and H. Emling for fruitful discussions.


\section*{References}
\begin{thebibliography}{99}
\bibitem{Mor}S. Mordechai {\it et al,} \Journal{\PRL}{60}{408}{1988}
\Journal{;}{61}{531}{1988}.

\bibitem{Rit}J. Ritman {\it et al,} \Journal{\PRL}{70}{533}{1993}.\\
             R. Schmidt {\it et al,} \Journal{\PRL}{70}{1767}{1993}.\\
             T. Aumann {\it et al,} \Journal{\PRC}{47}{1728}{1993}.

\bibitem{Eml}H. Emling, \Journal{\PPN}{33}{729}{1994}.\\
             Ph. Chomaz and N. Frascaria, \Journal{\PRP}{252}{275}{1995}.

\bibitem{Nis}S. Nishizaki and J. Wambach, \Journal{\PLB}{349}{7}{1995}.

\bibitem{Vey}S. Veyssiere {\it et al,} \Journal{\NPA}{159}{561}{1970}.

\bibitem{Ald}A. Winter and K. Alder, \Journal{\NPA}{319}{518}{1979}.\\
             K. Alder and A. Winter, {\em Coulomb Excitation},
             (Academic Press, New York and London, 1966).

\bibitem{Ber}C.A. Bertulani and G. Baur, \Journal{\PRP}{163}{299}{1988}.

\bibitem{Type}S. Typel and G. Baur, \Journal{\PRC}{50}{2104}{1994}.
 
\bibitem{Bor}K. Boretzky {it et al,} \Journal{\PLB}{384}{30}{1996}.

\bibitem{Pon}V.Yu. Ponomarev {\it et al.} \Journal{\PRL}{72}{1168}{1994}.

\end{thebibliography}
\end{document}